# Fully stabilized 25 GHz frequency comb for frequency calibration of optical spectrum analyzer

Yoonkwon On[1,2,6], Dae Hee Kim[1,6], Sujin Kim[1,3], Yong Jin Kim[1,4], Jong-Ahn Kim[1], Sunghoon Eom[1], Jae Yong Lee[1], Yoon-Soo Jang[1,2,5,*]

[1]Length and Dimensional Metrology Group, Division of Physical Metrology, Korea Research Institute of Standards and Science (KRISS), 267 Gajeong-ro, Yuseong-gu, Daejeon, 34113, Republic of Korea

[2] Department of Precision Measurement, University of Science and Technology (UST), Daejeon, 34113, Rep. of Korea

[3]Department of Physics, Chungnam National University, Daejeon, 34134, Rep. of Korea

[4]Department of Electronics Engineering, Chungnam National University, Daejeon, 34134, Rep. of Korea

[5]School of Electrical and Electronics Engineering, Chung-Ang University, Seoul, 06974, Rep. of Korea

[6]These authors are equally contributed

*ysj@kriss.re.kr

**Abstract**

Optical spectrometers are widely used in scientific and industrial applications, and precise frequency calibration is essential for ensuring their reliable performance. Traditionally, spectrometers have been calibrated using reference gas cells or reference lamps. However, such conventional methods are not enough to meet the demands for high accuracy and stability. Although frequency-stabilized lasers offer excellent frequency uncertainty, they provide only a single calibration point at a fixed frequency, which is unsuitable for wide-range spectrometer calibration. In this work, we demonstrate a fully stabilized, high-repetition rate electro-optic frequency comb (EO comb) as an absolute frequency reference providing multiple calibration points over a broad spectral range. Our 25 GHz EO comb provides a broadband spectrum spanning from 189.5 THz to 196 THz (corresponding to 1530 nm to 1582 nm), traceable to a frequency standard with a relative standard uncertainty of $10^{-13}$. The well-defined and evenly spaced comb modes can be spectrally resolved by conventional optical spectrometers, enabling wide-range, high-precision frequency calibration. We directly calibrated a commercial spectrometer by referencing its measured frequency values to our well-defined comb modes, thereby evaluating the frequency error with a standard uncertainty of 20 MHz (or relative standard uncertainty of $10^{-7}$), which is limited by the Type A uncertainty (repeatability) of the spectrometer. The proposed method is simple to implement and provides multiple calibration points referenced to the frequency standards over a broad spectral range. This approach improves both the calibration and performance evaluation of spectrometers, and it contributes to the advancement of optical metrology.

## Introduction

Spectrometers are widely used to characterize optical spectral properties ranging from science and industry [1], including astronomical science [2], optical communications [3], chemical analysis [4], biological analysis [5], and optical metrology [6]. Spectrometers typically rely on diffraction elements or Fourier transform techniques, and their frequency axis can be affected by a variety of error sources, including geometrical deviations of optical components, assembly and alignment error, thermal and mechanical instabilities, and detector noise. To compensate for these errors, frequency calibration is generally performed using reference lamps such as Neon, Krypton and Argon [7-11], reference gas cells [12-15] and low coherence interferometry [16,17]. Most commercial spectrometers exhibit frequency-axis error on the order of several GHz [18,19], which is insufficient to meet the demands for high accuracy measurements as science and technology advance [2,6,20]. Laser frequencies recommended by the CIPM (International Committee for Weights and Measures) offer a relative uncertainty at the $10^{-12}$ level [21], but as single-wavelength sources, they are not suitable for frequency calibration over the wide spectral range of spectrometers.

Frequency combs consist of a series of equally spaced laser modes in the frequency domain and serve as transfer oscillators that link optical frequencies to RF frequency standards, enabling relative uncertainty below $10^{-12}$ [22-24]. Such unique advantages of frequency combs have led to breakthroughs in optical metrology over the past two decades [25-28]. Frequency combs are typically realized using mode-locked femtosecond lasers [29-31]. However, the mode-spacing of most conventional mode-locked femtosecond lasers is limited to less than a few GHz due to limitation of the cavity length [32,33]. Since this mode-spacing is insufficient to be spectrally resolved by most commercial spectrometers, frequency combs based on the mode-locked femtosecond laser are generally unsuitable for frequency calibration of spectrometers. In space-based spectrometers, where the detection of extremely small frequency shift is required for exoplanet searches, high accuracy reference laser sources with wide spectral range and large mode-spacing are essential for precise frequency calibration [2,25]. To meet this requirement, the repetition rate multiplied frequency comb [34,35], alternative frequency comb with tens of GHz repetition rates such as microcomb [36,37] and electro-optic frequency comb (EO comb) [38,39] have been introduced and utilized to calibrate frequency of space-based spectrometers.

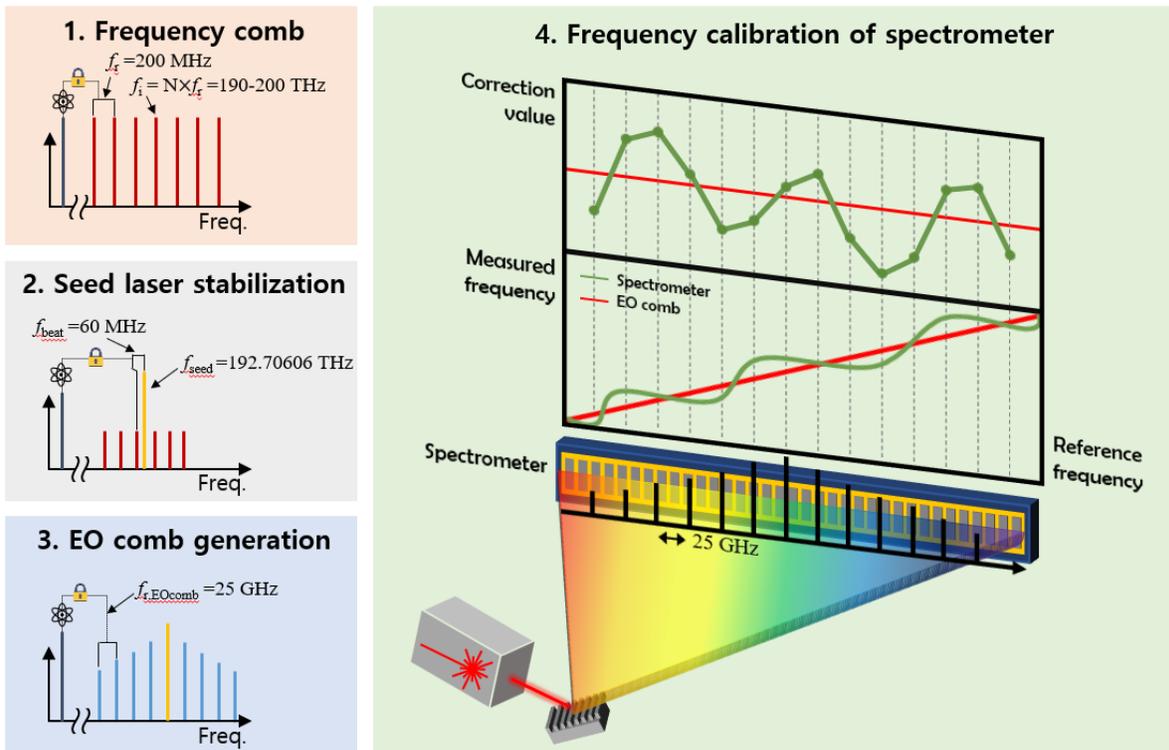

**Fig. 1.** Basic principle of EO comb-based calibration of spectrometer.

In particular, the EO combs generate frequency combs by producing sidebands through electro-optic modulation of a single wavelength laser [40-42]. Unlike mode-locked femtosecond laser or microcombs, the EO combs do not require an optical cavity. They offer convenient operation and construction, and allow flexible control of the repetition rate above 10 GHz with traceability to frequency standards. These features make the EO combs highly suitable for a wide range of applications, including optical metrology and optical communications [43-47].

In this work, we propose a frequency-calibration method for optical spectrometers using a fully stabilized, broadband 25 GHz EO comb. A single-wavelength laser, stabilized to a reference fiber frequency comb, is used as the seed laser [48,49]. A resonant EO comb generator subsequently produces a broadband 25 GHz EO comb with traceability to frequency standards [50-52]. The high repetition rate allows individual frequency comb modes to be spectrally resolved by conventional optical spectrometers. Additionally, the broadband spectrum provides numerous calibration points spanning from 189.5 THz to 196 THz (corresponding to 1530 nm to 1582 nm). The traceability to frequency standards, at the relative uncertainty level of $10^{-13}$, supports both high-precision and high-accuracy frequency calibration of spectrometers [53,54]. These advantages make our fully stabilized, broadband 25 GHz EO comb highly beneficial for the frequency calibration of optical spectrometers. As a feasibility demonstration, we evaluate the frequency-axis error of a commercial spectrometer using the proposed EO comb. The frequency values measured by the spectrometer are compared to the well-defined modes of EO comb, and calibration values are derived using a look-up table approach. While our EO comb provides calibration-point frequencies with a standard uncertainty at the 20 Hz level, the calibration result for the spectrometer exhibits a standard uncertainty of approximately 20 MHz, limited by the A-type uncertainty (repeatability) inherent to the spectrometer. The proposed method is straightforward, intuitive and easy to implement, and it provides broadband, frequency-traceable calibration points for optical spectrometers.

## Results

**Basic principle of frequency comb-based frequency calibration of spectrometer**

The frequency traceability scheme of the light source used for spectrometer calibration is shown in Fig. 2. The frequency standard is an atomic clock provided by the Time and Frequency Group at the Korea Research Institute of Standards and Science (KRISS). The 10 MHz Atomic clock signal has an expanded uncertainty (k=2) of $10^{-13}$ [55]. The reference frequency comb (DFC CORE +, Toptica comb) that is phase-locked to an atomic clock has a center wavelength of 1550 nm, a repetition rate ($f_r$) of 200 MHz and no carrier envelope offset frequency ($f_{ceo}$). Since the distributed feedback (DFB) laser serving as the seed for EO combs exhibits large frequency fluctuations in the free-running state, it needs to be stabilized to a single designated mode of the reference comb.

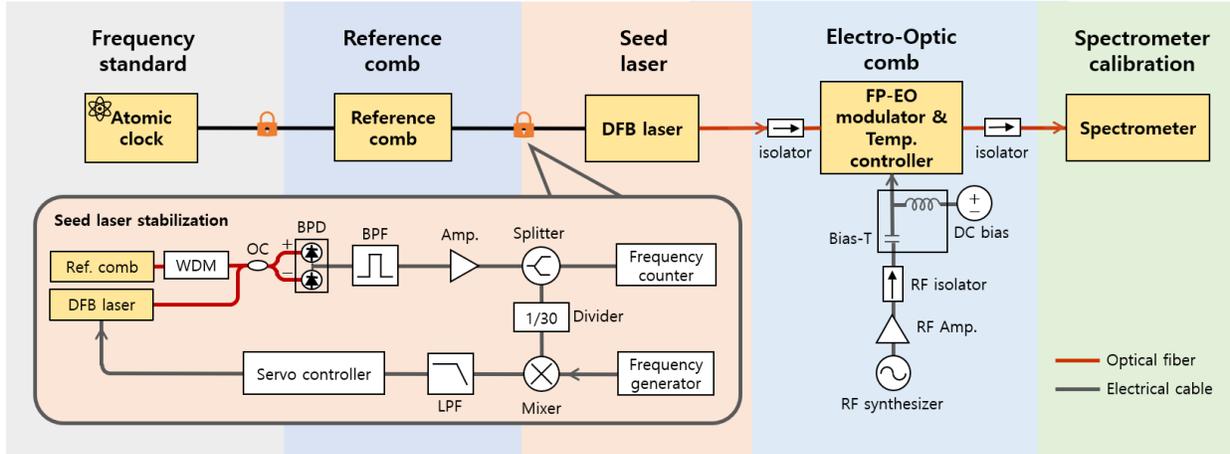

**Fig. 2.** Optical layout of traceable frequency comb-based calibration of spectrometer. WDM: wavelength division multiplexing, OC: optical coupler, BPD: balanced photodetector, BPF: band-pass filter, Amp: amplifier, LPF: low-pass filter.

The stabilization process is as follows. A C27 wavelength division multiplexing (WDM) filter is used to transmit the reference comb modes selectivity around 1555.75 nm, corresponding to the DFB laser's wavelength (1555 ± 10) nm. The filtered reference comb and DFB laser output are combined via an optical coupler. The generated beat signal is detected by a balanced photodetector (BPD), and passes through a low-pass filter (LPF) and a band-pass filter (BPF), extracting only the beat frequency ($f_{beat}$) required for stabilization. The beat frequency is divided by a factor of 30 for stable PLL control, allowing it to be phase-locked to a 2 MHz RF signal. The beat frequency is stabilized at 60 MHz (30 × 2 MHz), and the seed laser frequency ($f_{seed}$) is determined by the following equation,

$$f_{seed} = f_N \pm f_{beat} = N \cdot f_r \pm f_{beat} = N \cdot (200 \text{ MHz}) \pm 60 \text{ MHz} \quad (1)$$

Here, $N$ is the mode number of the reference comb mode that interferes with the DFB laser. The integer $N$ and the appropriate sign (±) are identified using a wavelength meter (Agilent 86122A, Agilent Technologies) with an accuracy of 0.3 pm (approximately 50 MHz). Eventually, the frequency of the seed laser becomes traceable to the atomic clock.

**Frequency stabilization of seed laser with $10^{-13}$ frequency uncertainty**

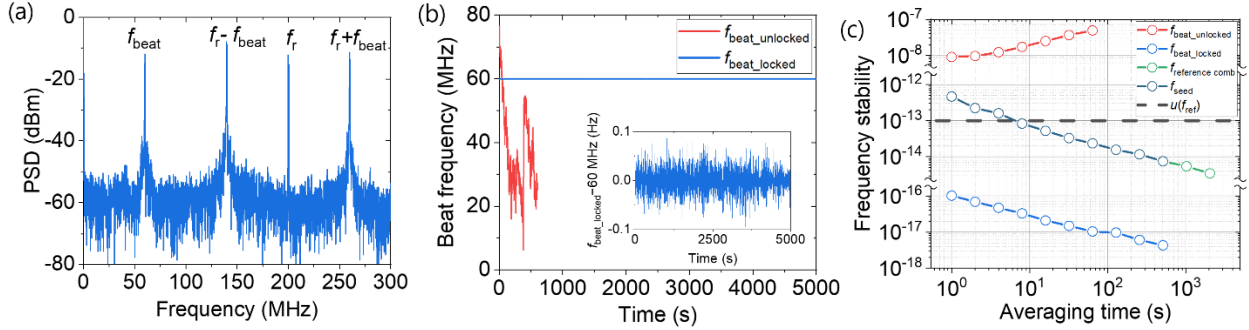

**Fig. 3.** Frequency stabilization of seed laser to reference frequency comb. (a) RF spectrum of the beat signal between the DFB laser and the reference comb. (b) Frequency fluctuations of the DFB laser beat signal. The inset shows the beat frequency of the DFB laser after locking, with a standard deviation of 0.02 Hz. (c) Frequency stability of the measured signals and the relative uncertainty of frequency standard. The stabilized DFB laser trace closely overlaps with that of the reference comb, confirming that the system achieves a stability level of $10^{-13}$.

Figure 3(a) shows the relationship (as described in equation (1)) between the stabilized DFB laser and the reference comb in the RF spectrum. A strong beat signal ($f_{beat}$) is observed at 60 MHz, along with the reference comb's repetition rate ($f_r$) of 200 MHz, as well as the sum and difference frequencies of 140 MHz ($f_r - f_{beat}$) and 260 MHz ($f_r + f_{beat}$), are all present. The performance of this stabilization is confirmed by observing the frequency fluctuations. As shown in Fig. 3(b), the DFB laser in the free-running state (red), which is not locked to a specific comb mode, exhibits large fluctuations exceeding 40 MHz. In contrast, the inset reveals that the laser stabilized to the reference comb (blue) is controlled with a standard deviation of only 0.02 Hz, maintaining the beat frequency precisely at 60 MHz. Based on the stable beat signal, the absolute frequency of the seed laser was determined by identifying the mode number ($N$) and its sign using a wavelength meter. The measurement confirmed the center frequency to be 192,707,060 MHz, which corresponds exactly to a frequency offset by 60 MHz from the reference comb mode at 192,707,000 MHz.

An analysis of the stabilization performance using the Allan deviation is shown in Fig. 3(c). The frequency stability improved from $10^{-8}$ in the free-running state to $10^{-13}$ after stabilization to the reference comb. This performance is limited by the relative uncertainty of frequency standard ($10^{-13}$) and represents an improvement of five orders of magnitude compared to the free-running state.

**Broadband and high-repetition rate frequency comb generation with $10^{-13}$ frequency uncertainty**

An electro-optic (EO) comb is generated by modulating a continuous-wave (CW) laser using an electro-optic modulator (EOM) driven by an external RF signal. In this method, the repetition rate ($f_{r,EOcomb}$) is determined by the externally driving RF frequency, facilitating the generation of high repetition rate combs. Thus, the resulting large mode spacing allows individual comb modes to be resolved using spectrometers.

In this study, considering the thermal sensitivity of the cavity length, we implemented an EO comb using a temperature-controlled resonant generator (WTEC-02-25, OptoComb). The device features an EOM positioned within a Fabry-Pérot cavity with a 2.5 GHz free spectral range (FSR). A stable frequency comb was generated by applying a 25 GHz RF signal, matching the 10th harmonic of the FSR. The generated comb exhibits a mode spacing of 25 GHz, seeded by a 192.70706 THz laser with a relative uncertainty of $10^{-13}$. The frequency of the $i$-th EO comb mode, $f_i$, can be expressed by the equation:

$$f_i = f_{seed} \pm i \cdot f_{r,EOcomb} = (192.70606 \text{ THz}) \pm i \cdot (25 \text{ GHz}) \tag{2}$$

Figure 4(a) shows the spectrum of the broadband EO comb measured using a commercial spectrometer with a resolution of 4 pm (~500 MHz), where individual comb modes are resolved. The observed triangular spectral envelope is characteristic of the resonant nature of the EO comb [50]. The generated comb lines, spaced by 25 GHz, span a broad bandwidth from 189.5 THz to 196 THz (covering the C-band and part of the L-band), centered at the 192.70706 THz seed laser.

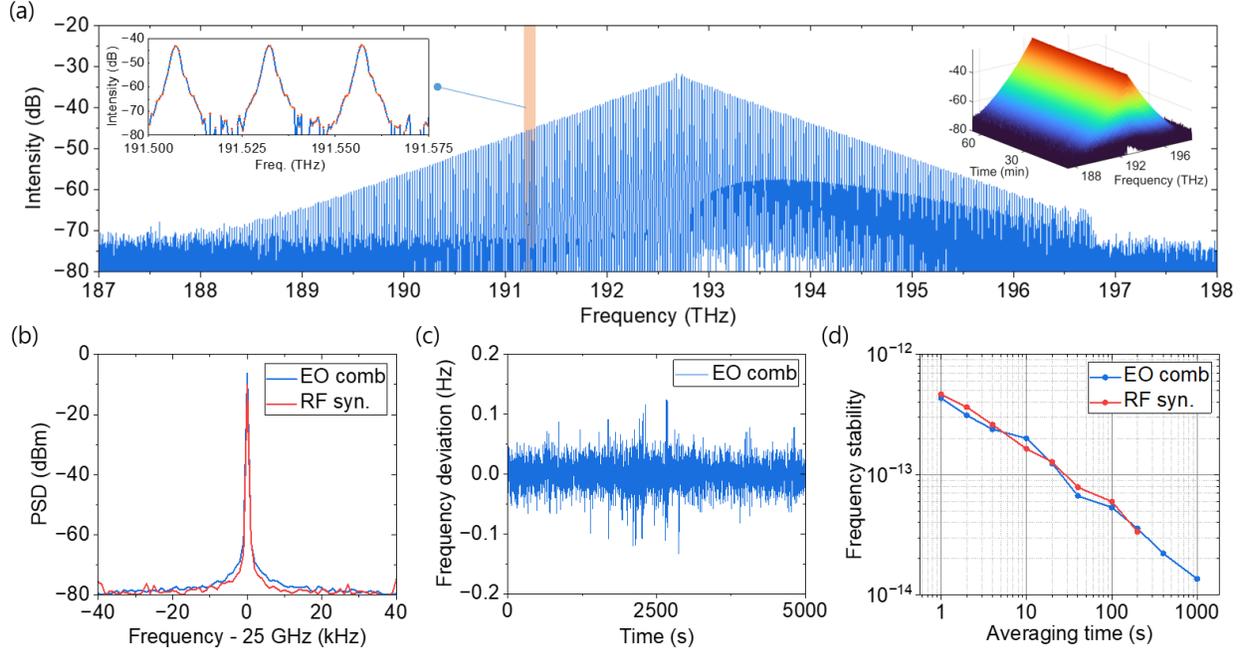

**Fig. 4.** Generation and characterization of the EO comb. (a) Optical spectrum of the EO comb with a 25 GHz repetition rate. Insets show three adjacent modes with a 25 GHz mode spacing (top-left) and the 3D spectral stability measured over time (right). (b) RF power spectral density (PSD) of the driving synthesizer and the generated EO comb signal. (c) Time-domain frequency deviation of the EO comb. (d) Frequency stability of the RF synthesizer and the EO comb.

High-precision EO comb generation necessitates high stability in both the seed laser and the RF modulation signal. Since the seed laser's performance was confirmed in the previous section, we proceeded to characterize the 25 GHz RF signal. Fig. 4(b) presents the RF spectrum with a signal-to-noise ratio (SNR) exceeding 60 dB, confirming that the EO comb closely matches the high signal quality of the synthesizer. To analyze long-term stability, the EO comb's repetition rate was recorded for 5,000 seconds (Fig. 4(c)). The resulting Allan deviation plotted in Fig. 4(d) verifies that both the synthesizer and the generated EO comb satisfy the stability level of $10^{-13}$.

**Frequency calibration of spectrometer using EO comb**

We utilized the stably generated Electro-Optic (EO) frequency comb to calibrate the frequency axis of a commercial spectrometer. The comb's center frequency was measured at 192.7076 THz by the spectrometer. The EO comb features modes spaced by 25 GHz relative to the center frequency, with each mode exhibiting a relative uncertainty of $10^{-13}$. Consequently, 248 comb modes with precisely known absolute frequencies are available as reference points across the range from 189.5 THz to 196 THz. Comparing the absolute frequencies of these comb modes with the spectrometer's measurements enables precise calibration across the entire frequency axis.

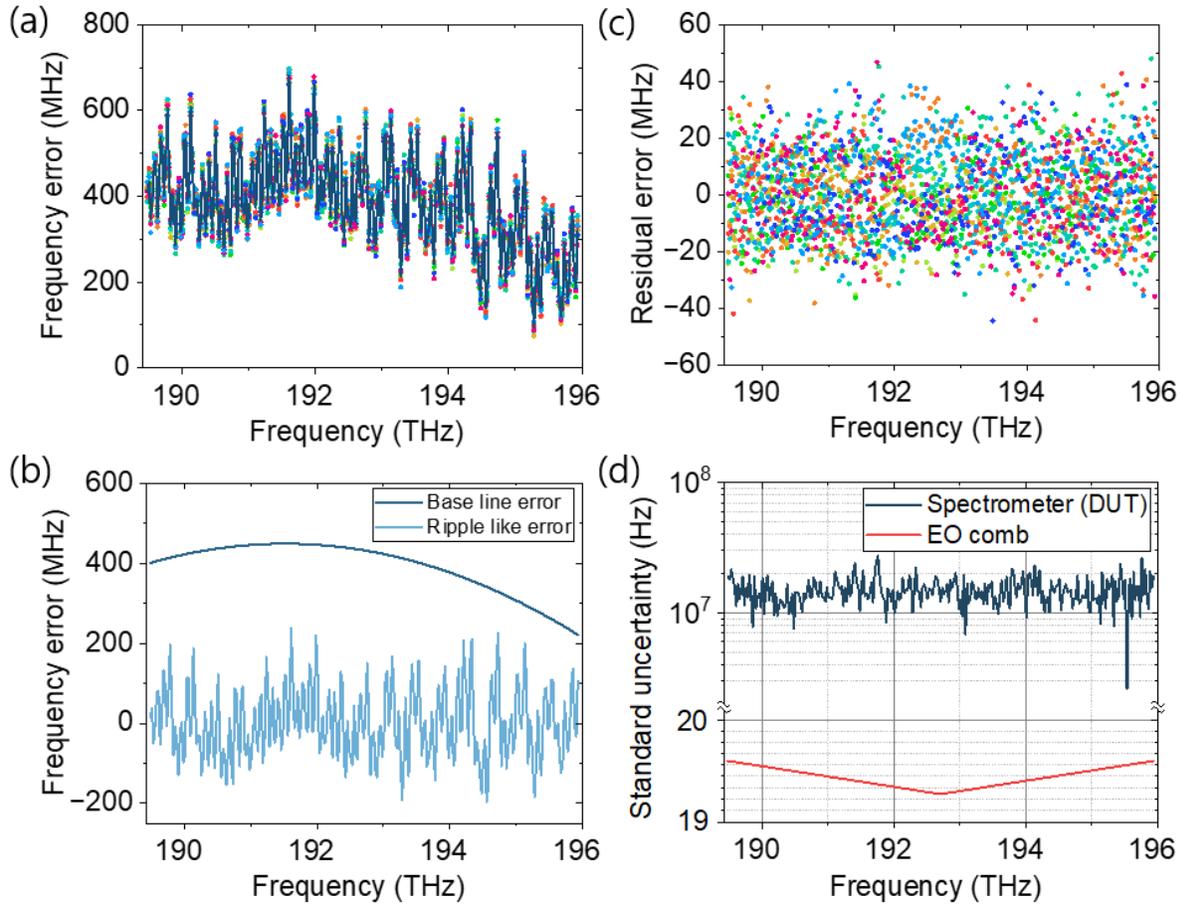

**Fig. 5.** Analysis of frequency errors and calibration performance. (a) Measured frequency errors for 10 repeated measurements, overlaid with the calibration curve derived from these data points. (b) Decomposition of the calibration curve into baseline- and ripple-like error components. These correspond to the global trend correctable by conventional methods and local fluctuations requiring dense calibration points, respectively. (c) Residual frequency error after applying the LUT correction, showing a reduction to 20 MHz (spectrometer resolution limit). (d) Comparison of standard uncertainties: the experimental limit imposed by the spectrometer (~20 MHz) versus the intrinsic uncertainty of the proposed EO comb method (20 Hz, calculated via Eq. (3)).

We conducted 10 repeated measurements and derived a calibration curve based on the average error profile. These results are shown in Fig. 5(a), where the calibration curve is overlaid on the measured data points. Using this dataset, we decomposed the error characteristics into low- and high-frequency components (Fig. 5(b)). While the conventional 2nd-order polynomial fit addresses only the global baseline trend, our look-up table (LUT) method utilizes the dense comb modes to correct local ripple components. Consequently, the polynomial fit leaves residual errors of approximately 200 MHz, whereas the LUT correction reduces the error to 20 MHz (Fig. 5(c)), limited by the A-type uncertainty (repeatability) inherent to the spectrometer. Finally, Fig. 5(d) compares the spectrometer-limited uncertainty (~20 MHz) with the intrinsic uncertainty of our method, expressed as 20 Hz in Eq. (3):

$$u(f_i) = u(f_{seed}) \pm i \cdot u(f_{r,EOcomb}) = (192.70606 \text{ THz} \cdot 10^{-13}) \pm i \cdot (25 \text{ GHz} \cdot 10^{-13}) \qquad (3)$$

where $i$ is the mode number of EO comb. This significant gap confirms that the current calibration uncertainty is dominated by the spectrometer's repeatability, while the proposed method maintains a significantly lower intrinsic uncertainty of 20 Hz.

**Conclusion**

In this study, we presented a precision calibration method using an EO comb as a traceable 'optical frequency ruler.' We corrected the spectrometer's frequency axis using comb modes with an uncertainty of $10^{-13}$ across the 189.5 - 196 THz range. Consequently, the calibrated frequency axis achieved a standard uncertainty of 20 MHz, a limit imposed by the spectrometer's Type A uncertainty. However, our proposed reference method maintains an intrinsic uncertainty of only 20 Hz ($10^{-13}$), offering a calibration precision far exceeding the limitations of current commercial spectrometers.

Our proposed method is advantageous because it is simple to implement and provides numerous calibration points traceable to a frequency standard across a broad bandwidth. Building on these strengths, this approach can be readily extended beyond spectrometer calibration to the field of wavelength meter calibration. Furthermore, it holds significant potential for application in astronomical spectroscopy as an 'astro-comb,' a crucial tool for discovering exoplanets. Ultimately, this research is expected to contribute to the advancement of precision optical metrology.

**Acknowledgments**: This work is supported by Korea Research Institute of Standard and Science (25011026, 25011211).

**Author contributions:** Y.-S. J. led the project and designed the experiment. Y.-S. J. and Y. O. performed the electro-optic frequency comb generation. Y.-S. J., D. H. K. and S. E. performed the frequency stabilization of the reference frequency comb and seed laser. Y. O., D. H. K., S. K., Y. J. K., J.-A. K., and J. Y. L. performed the frequency calibration of optical spectrum analyzer. Y. O., D. H. K., S. K., and Y. J. K. conducted the analysis of the measured data. All authors prepared the manuscript.

**Data availability**: The data that support the plots within this paper and other finding of this study are available from the corresponding author upon reasonable request.

# Supplementary material

# Fully stabilized 25 GHz frequency comb for frequency calibration of optical spectrometers

Yoonkwon On[1,2,6], Dae Hee Kim[1,6], Sujin Kim[1,3], Yong Jin Kim[1,4], Jong-Ahn Kim[1], Sunghoon Eom[1], Jae Yong Lee[1], Yoon-Soo Jang[1,2,5,*]

[1]Length and Dimensional Metrology Group, Division of Physical Metrology, Korea Research Institute of Standards and Science (KRISS), 267 Gajeong-ro, Yuseong-gu, Daejeon, 34113, Republic of Korea

[2] Department of Precision Measurement, University of Science and Technology (UST), Daejeon, 34113, Rep. of Korea

[3]Department of Physics, Chungnam National University, Daejeon, 34134, Rep. of Korea

[4]Department of Electronics Engineering, Chungnam National University, Daejeon, 34134, Rep. of Korea

[5]School of Electrical and Electronics Engineering, Chung-Ang University, Seoul, 06974, Rep. of Korea

[6]These authors are equally contributed

*ysj@kriss.re.kr

## 1. Appendix A: Long-term Spectrum and Relative intensity noise (RIN) Analysis

To verify the operational robustness of the system, we evaluated the long-term stability of the optical frequency comb. Fig. S1(a) displays the optical spectra recorded continuously over a period of 1 hour. The superimposed spectral traces exhibit negligible deviation, demonstrating highly consistent spectral performance over time. In addition to spectral stability, we characterized the amplitude noise of the comb source. As shown in Fig. S1(b), the Relative Intensity Noise (RIN) was measured to assess the noise properties. The result indicates a highly stable output, with the RIN floor dropping to as low as -140 dBc/Hz. These results confirm that the proposed 25 GHz comb maintains high stability in both frequency and amplitude domains, making it suitable for precision metrology applications.

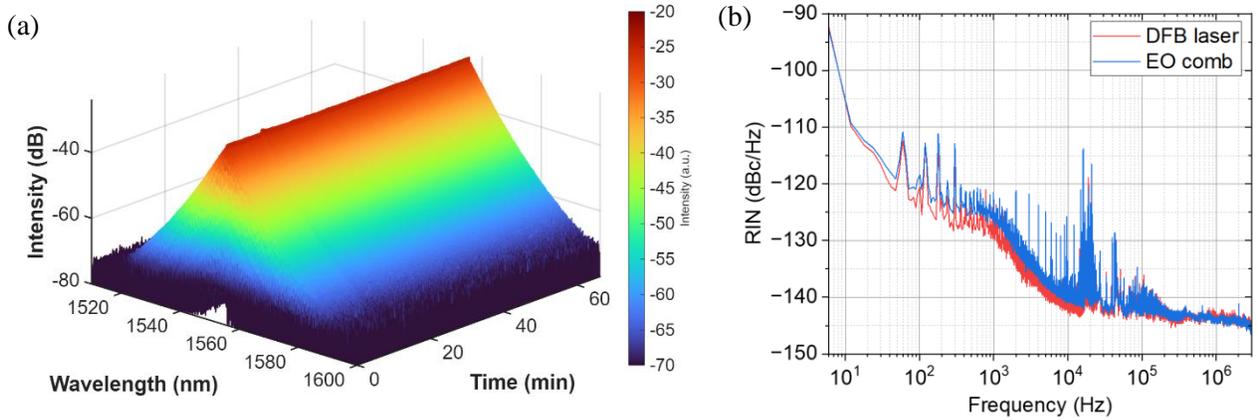

**Fig. S1.** Long-term stability and noise characterization of the 25 GHz EO comb. (a) Optical spectra measured continuously for 1 hour, showing excellent long-term spectral stability. (b) Relative Intensity Noise (RIN) measurement result. The noise level drops to -140 dBc/Hz, indicating the low-noise characteristics of the stabilized comb.

## 2. Appendix B: Peak detection using polynomial fitting

To achieve MHz-level resolution, a method to find the peak of the frequency ($f_{peak}$) is required. This was accomplished by applying a second-order polynomial fit to the peak signal, creating a continuous model of the form $I(f) = Af^2 + Bf + C$. The position of the peak maximum corresponds to the function's extremum, which is found where its first derivative, $dI/df$, equals zero. Solving $2Af + B = 0$ provides the precise peak location at $f = -B/2A$. This fitting method provides a significant precision improvement over direct peak identification.

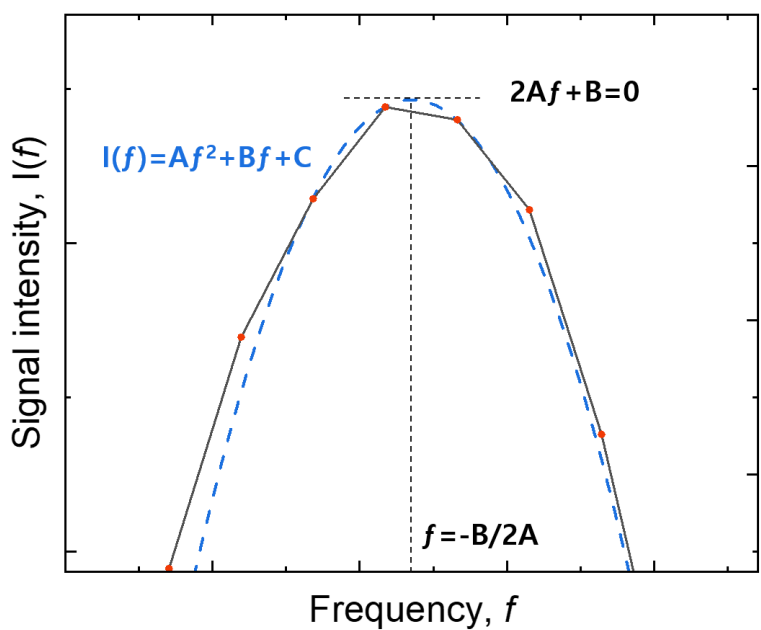

**Fig. S2.** Precise peak detection by 3-point second-order polynomial curve fitting

# 3. Appendix C: Enhancement of Calibration Precision via Sub-pixel Peak Detection

Given the system's sampling resolution of approximately 400 MHz per pixel, the conventional Peak-point method is inherently limited by quantization errors, as it is bound to the discrete sampling grid. In contrast, the Poly-fit method circumvents this hardware limitation as illustrated in Fig. S2(a). By fitting the raw spectral data (red dots) to a local quadratic curve, the precise peak location (blue dot) is identified within the sub-pixel domain. The efficacy of this approach is demonstrated in Fig. S2(b), which compares the frequency calibration errors of the two methods. The Poly-fit method (blue) exhibits significantly improved stability compared to the discrete Peak-point detection (red). Specifically, after removing the systematic offset, the maximum frequency error is reduced from approximately 400 MHz to within 200 MHz, confirming that the proposed approach effectively compensates for the coarse sampling interval.

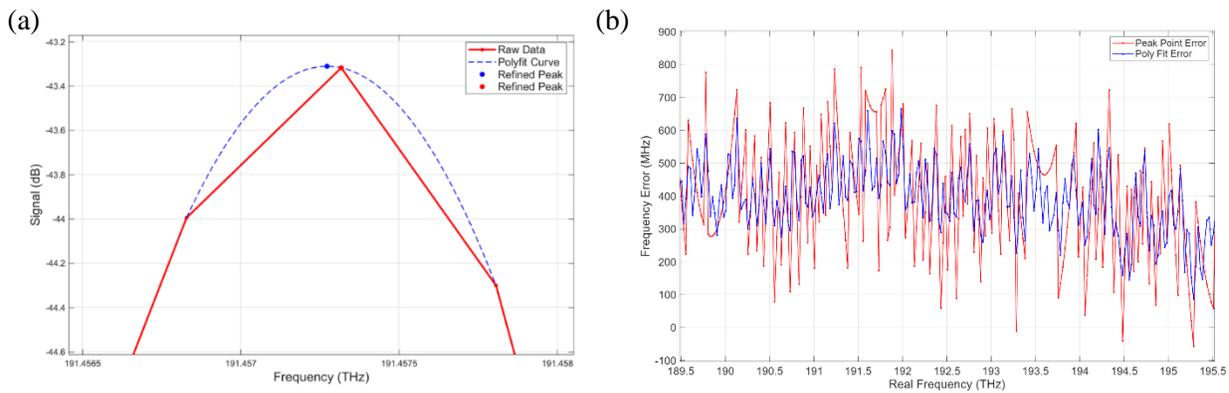

**Fig. S3.** Comparison of peak detection methods. (a) Sub-pixel peak estimation using local quadratic fitting. The red dots indicate raw spectral data, and the blue dot represents the fitted peak. (b) Frequency calibration errors for the discrete Peak-point method (red) and the proposed Poly-fit method (blue).

# 4. Appendix D: Enhancement of Calibration Precision via Sub-pixel Peak Detection

Table S1 presents a performance comparison with previous works on frequency calibration of optical spectrometers.

Table S1. Summary comparison on frequency calibration of optical spectrometers

| Method | Repetition rate | Wavelength range | Frequency uncertainty | Reference |
|---|---|---|---|---|
| Reference lamp | N/A | 500-1000 nm | $1 \cdot 10^{-4}$ | [7] |
| Absorption cell | N/A | 1520-1560 nm | $1 \cdot 10^{-7} - 10^{-8}$ | [14] |
| Low coherence interferometry | N/A | 500-900 nm | $1 \cdot 10^{-4}$ | [16] |
| EO comb | 12 GHz | 1430-1640 nm | $1 \cdot 10^{-9}$ | [38] |
| Filtered fiber comb | 25-37 GHz | 820-920 nm | $1 \cdot 10^{-12}$ | [2] |
| Microcomb | 22 GHz | 1471-1731 nm | $5 \cdot 10^{-9}$ | [36] |
| Microcomb | 23.7 GHz | 1486-1685 nm | $1 \cdot 10^{-12}$ (GPS/Rb clk) | [37] |
| **Resonant EO comb** | **25 GHz** | **1520 nm – 1580 nm** | **$10^{-13}$** | **This work** |

**References**

S1. Park, Min-Cheol, and Seung-Woo Kim. "Direct quadratic polynomial fitting for fringe peak detection of white light scanning interferograms." Optical Engineering 39.4 (2000): 952-959.